\documentstyle[12pt]{article}
\def\lsim{\mathrel{\rlap {\raise.5ex\hbox{$ < $}}
{\lower.5ex\hbox{$\sim$}}}}

\newcommand{\pr}{\paragraph{}}
\newcommand{\be}{\begin{equation}}
\newcommand{\ee}{\end{equation}}
\newcommand{\bea}{\begin{eqnarray}}
\newcommand{\nn}{\nonumber}
\newcommand{\eea}{\end{eqnarray}}

\newcommand{\nk}{\noindent}
\def\gappeq{\mathrel{\rlap {\raise.5ex\hbox{$>$}}
{\lower.5ex\hbox{$\sim$}}}}

\def\lappeq{\mathrel{\rlap{\raise.5ex\hbox{$<$}}
{\lower.5ex\hbox{$\sim$}}}}

\begin{document}

\begin{titlepage}
\begin{flushright}
ACT-03/95 \\
CERN-TH.7480/94 \\
CTP-TAMU-12/95 \\
ENSLAPP-A.492/94 \\
\end{flushright}
\begin{centering}
\vspace{.1in}
{\large {\bf A String Scenario for Inflationary Cosmology }} \\
\vspace{.2in}
{\bf John Ellis}$^{a}$,
{\bf N.E. Mavromatos}$^{b,\diamond}$ and
{\bf D.V. Nanopoulos}$^{a,c}$
\\
\vspace{.03in}
\vspace{.1in}
{\bf Abstract} \\
\vspace{.05in}
\end{centering}
{We describe a scenario for inflation in the framework of
non-critical string theory,
which does not employ an inflaton field.
There is an exponential expansion of the
volume of the Universe, induced by enormous
entropy production in the
early stages of cosmological
evolution. This is associated with
the loss of information
carried by global string modes that cross the
particle horizon.
It is the same loss of information that induces irreversible
time flow when target
time is identified
with the world-sheet Liouville mode. The resulting
scenario for
inflation is described by a string analogue of the
Fokker-Planck equation that
incorporates
diffusion and dissipative effects.
Cosmological density perturbations are naturally
small.}

\vspace{0.2in}
\nk $^a$ Theory Division, CERN, CH-1211, Geneva, Switzerland,  \\
$^{b}$ Laboratoire de Physique Th\`eorique
ENSLAPP (URA 14-36 du CNRS, associe\`e \`a l' E.N.S
de Lyon, et au LAPP (IN2P3-CNRS) d'Annecy-le-Vieux),
Chemin de Bellevue, BP 110, F-74941 Annecy-le-Vieux
Cedex, France ;\\
$^{\diamond}$ On
leave from P.P.A.R.C. Advanced Fellowship, Dept. of Physics
(Theoretical Physics), University of Oxford, 1 Keble Road,
Oxford OX1 3NP, U.K.  \\
$^{c}$ Center for
Theoretical Physics, Dept. of Physics,
Texas A \& M University, College Station, TX 77843-4242, USA
and Astroparticle Physics Group, Houston
Advanced Research Center (HARC), The Mitchell Campus,
Woodlands, TX 77381, USA. \\
\vspace{0.01in}
\begin{flushleft}
ACT-03/95 \\
CERN-TH.7480/94 \\
CTP-TAMU-12/95 \\
ENSLAPP-A.492/94 \\

March 1995 \\
\end{flushleft}
\end{titlepage}
\newpage
\section{Introduction}
\pr
Inflation\cite{guth,vilenkin}
 offers an attractive solution to many of the major
problems of conventional Big Bang cosmology, such as the age, size and
homogeneity of the observed Universe, its large
entropy and its closeness to the critical density.
Inflation occurs for generic scalar field potentials,
but these may not fit naturally within a conventional
Grand Unified field theory. A number of proposals
have been made for the accommodation of inflation
within string theory\cite{enqo,inflstring}, but
these are also beset with difficulties.
At the level of an effective field theory derived from an
underlying string theory, it becomes even more difficult
to generate naturally a suitable inflationary
scalar potential.
Some such scenaria
have inflationary epochs that are too short and/or
inhomogeneous density
fluctuations that are too large\cite{enqo}.
However, we do not rule out the possibility that a realistic
scenario may emerge under certain extra assumptions.
\pr
A more fundamental approach has been to study full
string theories in time-dependent backgrounds\cite{aben}, in the hope
of identifying string solutions of cosmological
interest that could serve as the basis of an inflationary
model. Consistent partition functions for such
cosmological string theories have been constructed,
featuring a dilaton field that depends linearly on time.
This is linked to a deficit $Q^2$ in the central charge
that is compensated by other degrees of freedom to
restore criticality. Matter fields move in such a
non-critical time-dependent background according
to equations of motion containing
extra `frictional' terms $\propto Q$, which plays the r\^ole of the
Hubble expansion parameter, and the time-dependent
dilaton can be identified as a non-critical Liouville
field, as we discuss in more detail later. Such
cosmological string models do not exhibit exponential
inflation, when treated at the classical level\cite{aben},
but we recall that the quantum effects play essential
r\^oles in conventional field-theoretical
inflation\cite{guth,vilenkin},
e.g. in providing initial fluctuations and in
converting scalar field energy into
matter-particle entropy\cite{abott}. We are therefore led to seek
a consistent
quantum treatment of time-dependent string theories,
which provides the string scenario for inflation
discussed below.
\pr
We first review salient features of conventional
field-theoretical
inflation, emphasizing a frictional interpretation of the
Hubble expansion and the relation of entropy generation
to the integration over field degrees of freedom
that disappear beyond the de Sitter horizon.
Then we construct a string analogue starting from the
Zamolodchikov function interpreted
as an effective action in string theory space\cite{zam,mm2},
demonstrating
that it leads to a consistent quantum formulation that
incorporates dissipation, associated
with couplings to unobserved string modes. The
previous cosmological
string theories \cite{aben} can be regarded as specific examples of this
general framework. In the cosmological
context, the dissipation is related
to {\it global} string modes \cite{emnerice} that
cannot be detected via {\it local} experiments
within a horizon volume. Then
we derive a string analogue of the
Fokker-Planck equation, demonstrate
that it may lead to the classical `ball rolling down the hill'
picture of conventional field theoretical
inflation\cite{guth,vilenkin},
as well as exponential increases in the cosmological
scale factor and matter-particle entropy\cite{abott}.
This inflation is achieved
without an inflaton field and can be
regarded as a world-sheet analogue of
Starobinski's ideas \cite{staro}.
The magnitude of density perturbations
is related to the ratio of particle and higher string mode energies,
and is naturally small.
\pr
\section{Review of Field-Theoretical Inflation}
\pr
In conventional field-theoretical inflation\cite{guth,vilenkin},
the (near-) exponential expansion of the Universe
is driven by the potential energy $V(\Phi )$
of a specially-introduced inflaton field
$\Phi$, treated classically $\Phi
 \simeq \Phi _c$ as a first approximation,
 evolving according to the
equation
\be
\nabla ^2 \Phi _c + \frac{\delta}{\delta \Phi _c}
V(\Phi ) | _{\Phi = \Phi _c}  =0
\label{ena}
\ee
Assuming spatial homogeneity,
inserting a Robertson-Walker-Friedmann scale factor
of $e^{Ht}$ into the first term in (\ref{ena}),
and approximating the second term by $-\mu^2 \Phi _c$,
one obtains
\be
   {\ddot \Phi _c} + 3 H {\dot \Phi _c} - \mu ^2
\Phi _c = 0
\label{dyo}
\ee
This equation exhibits frictional dissipation $\propto H$.
In the interesting limit of a slow rollover, the growing
solution of (\ref{dyo}) has
\be
\Phi _c \simeq exp (\frac{\mu ^2}{3H} t)
\label{tria}
\ee
yielding a typical rollover time
$ \tau \simeq \frac{3 H}{\mu ^2} $
and an exponential growth in the scale
factor by an amount
\be
R/R_0 \simeq exp (H\tau ) \simeq
e^{\frac{3 H^2}{\mu ^2}}
\label{pente}
\ee
This is sufficiently large ($H\tau \ge 60 $)
if the effective potential $V(\Phi _c )$ is
sufficiently flat, i.e.
\be
  3 H^2 \simeq 8\pi G_N V(\Phi _c ) \ge
60 \mu ^2 \simeq 60 V'' (\Phi _c )
\label{exi}
\ee
An order-of-magnitude lower bound on $\mu ^2 $ is provided
by the Hawking temperature
$ T_H =\frac{H}{2 \pi} $
associated with the presence of a horizon during the
de-Sitter-like inflationary epoch. This limits the
roll-over-time $\tau \le 3H/T_H^2 $, so that
the scale factor factor grows by a factor of at most
$exp (12 \pi ^2 )$, which is ample to accommodate our observable
Universe.
\pr
The finite temperature $T_H $
is related to the area
$  A = 12\pi / H^2 $
of the de Sitter horizon, in close parallelism to the
Hawking temperature of a black hole horizon. In each case,
information is lost across the horizon, as a
consequence of the necessary integration over unseen
field modes beyond the horizon. In the inflationary case,
the entropy per horizon volume $S_H \propto A$,
and the total entropy grows exponentially along with the scale
factor (\ref{pente}).
This can be regarded as due to the separation of the field
modes into two categories, only one of which, $\Phi _{<}$,
is inside the horizon and hence
observable, whilst the other acts as a
stochastic noise source in an application of the
fluctuation-dissipation theorem to the observable modes.
A stochastic framework has been suggested for inflation,
which is based on the following Fokker-Planck
equation
for the probability distribution ${\cal P} (\Phi _{<},t)$
of the observable modes
\cite{vilenkin,linde}
\be
\partial_t {\cal P}(\Phi_{<}, t) =
\frac{1}{8\pi^2}
\frac{\delta}{\delta \Phi _{<}} (
H^{\frac{3}{2}}
\frac{\delta}{\delta \Phi_{<}}[H^{\frac{3}{2}}
{\cal P} (\Phi_{<}, t)])
+ \frac{1}{6}\frac{\delta }{\delta \Phi_{<}}
[\frac{1}{H} \frac{\delta }{\delta \Phi_{<}}V (\Phi_{<})
{\cal P} (\Phi_{<}, t)]
\label{FPl}
\ee
where the standard deviation of the observable modes is
identified with
the zero-point fluctuation
of a massless scalar field in a de Sitter background,
\be
    <(\delta \Phi _{<} )^2>=\frac{1}{4\pi^2}H^3\delta t
\label{deviation}
\ee
Such
density
fluctuations in the observable Universe may be regarded
as
thermal fluctuations
associated with the finite horizon.
\pr
In the above
brief review, we have glossed over the transition
from the initial quantum regime to the classical
`ball rolling down the hill' picture (\ref{ena}), and it
is desirable to establish a unified description
of this, entropy generation during inflationary
epoch and the conversion into matter particles
at the end of inflation. Given the
progress in this direction
made in conventional field-theoretic
inflation\cite{linde}, we now seek a corresponding
framework in string theory.
\pr
\section{String Framework}
\pr
Our starting point is the recognition
that a suitable variant of the Zamolodchikov\cite{zam}
$C$ function $C[g^i ]$ can be regarded
as an effective action in the space
of two-dimensional field theories on the world-sheet,
parametrized by the couplings $g^i$ of
generalized $\sigma$-models
\bea
  C[g] &=& \int d^2 z (2 z^4
  <T_{zz} (z, {\bar z}) T_{zz} (0,0)>_{g}
- 3z^3{\bar z} <T_{zz} (z, {\bar z} ) T_{z{\bar z}} (0,0) >_{g} -
\nn \\
&~& 6 z^2 {\bar z}^2 <T_{z{\bar z}} (z, {\bar z})\Theta
(0,0) >_{g} )
\label{ennea}
\eea
where $T_{\alpha\beta}$ are components of the
world-sheet stress tensor, whose trace is $\Theta $.
The couplings $g ^i $ correspond in general to
non-conformal, relevant deformations $V_i$, with anomalous
dimensions $h _i < 2$ and non-trivial flow under the
renormalization group, which is described by
\be
    \beta ^i \equiv \frac{d }{d ln \mu } g^i
     = (h_i -2 )g^i -\pi c^i_{jk}g^jg^k
\label{deka}
\ee
where $\mu $ is a renormalization group scale and
the $c^i_{jk}$ are operator-product-expansion
coefficients for the $V_i$.
The corresponding classical equations of motion
are the familiar gradient flow
\be
     G_{ij} \beta ^j = \frac{\delta C}{\delta g^i}
\label{dodeka}
\ee
where $G_{ij} = <V_i V_j >$ is the metric in coupling
space.
\pr
Coupling (\ref{ennea}) to two-dimensional quantum gravity
restores\cite{aben,DDK} conformal invariance at the quantum
level, by dressing the vertex operators :
$V_i \rightarrow [ V_i ]_{\phi} $, where $\phi $ is the
Liouville field that scales the world-sheet metric
$
\gamma _{\alpha\beta}
= e^{\phi (z, {\bar z}) }
{\hat \gamma} _{\alpha\beta} $
where ${\hat \gamma }$ denotes a fixed fiducial metric.
In our approach, $\phi (z, {\bar z})$ serves as a local
renormalization scale on the world sheet.
The dressed vertex operators $[V_i]_{\phi }$ are then
{\it exactly}
marginal and the corresponding gravitationally-renormalized
couplings are \cite{DDK, schmid}
\be
   \lambda ^i (\phi ) = g^i e^{\alpha ^i \phi } +
\frac{\pi } {Q \pm 2 \alpha ^i } c^i_{jk} g^jg^k \phi
e^{\alpha _i \phi } + \dots
\label{dekatessera}
\ee
where \cite{DDK}
\be
Q  =  \sqrt{\frac{|25 - c|}{3} + \frac{1}{4}\beta ^i G_{ij} \beta ^j
}  \qquad ; \qquad
\alpha _i ^2  + \alpha _i Q  =
sgn(25 - c)(h_i - 2)
\label{dekapente}
\ee
is the central charge deficit
of the matter theory defined by
$g^i$, and the
appropriate gravitationally-dressed version of (\ref{ennea})
is supplemented by the Liouville action
\bea
   S_{mL} &=& S_m + S_L \nn \\
     S_L &=&
   \frac{1}{4\pi \alpha '} \int d^2z \sqrt{{\hat \gamma}}
[ (\partial _\alpha  \phi  )
(\partial ^\alpha \phi)
 -  Q \phi R^{(2)} ] \nn \\
S_m &=& \int d^2z \sqrt{{\hat \gamma}} \lambda ^i (\phi )  V_i
\label{dekaexi}
\eea
where $R^{(2)}$ is the curvature
of the world-sheet.
Note that we normalize the Liouville field so
as to have a canonical kinetic term on the world
sheet, enabling us to identify $\phi$ with the
target time, which we denote by $t$. This
will be the
physical time, observed in standard units.
However, for reasons that will become clear later on,
we introduce a rescaled time $\tau $:
\be
\phi \equiv t = Q\tau
\label{resc}
\ee
We shall come back to the physical time $\phi = t$
when we discuss our estimates for physical
observables.
The matter system is supercritical if $c > 25$, a possibility
first studied as a prototype for an expanding Universe in the
framework of string theory in ref. \cite{aben}, whose
`physical time' is
$e^{Q\phi}$ in our notation.
Equation (\ref{dekapente}) has the
solutions
\be
   \alpha _i^{\pm}  = -\frac{Q}{2} \pm \sqrt{Q^2 - (h_i -2)}  \qquad ;
\qquad c \ge 25
\label{dekaepta}
\ee
The branch of solutions $\alpha _i ^{-}$  corresponds to non-existing
operators in Liouville theory with the wrong boundary
conditions \cite{seib},
and should be excluded. This
leaves us with the physically-relevant solutions $\alpha _i^{(+)}$,
which vanish for $(1,1)$ operators, which are made exactly marginal
by the linear factor in the second term of (\ref{dekatessera}).
\pr
In ref. \cite{emnerice} we went further than in \cite{aben},
identifying the Liouville field $\phi = Q\tau $ with a world-sheet
renormalization scale via the world-sheet area
\be
A \equiv \int d^2z \sqrt{{\hat \gamma}}e^{\alpha \phi (z,{\bar z})}
\qquad ; \qquad
\alpha =-\frac{Q}{2} +
\frac{1}{2} \sqrt{ Q^2 \pm 8}
\label{dekaokto}
\ee
Equation (\ref{resc}) implies that derivatives with respect
to our target time
variable $t$
of the Liouville-dressed $(1,1)$ couplings in the neighborhood
of a fixed point are simply the conventional
$\beta$ functions (\ref{deka}) with the replacements
$g ^i \rightarrow \lambda ^i$:
\be
{\dot \lambda} ^i \equiv
\frac{d \lambda ^i }{d \tau} = \beta ^i (\lambda )
\label{eikosiena}
\ee
which makes a Liouville flow similar (formally) to a
conventional renormalization group flow on the world sheet.
Here and subsequently, derivatives with respect to
$\tau$ are denoted by dots.
In this formalism, the deficit $Q$, which is a constant at lowest
order in the coupling, becomes (\ref{dekapente})
a `running' central charge
that approaches a constant at a fixed point.
This renormalization group flow may be interpreted
as a friction problem \cite{emnerice}
with dissipation coefficient
$Q$, since $(1,1)$ deformations $\lambda ^i (\tau)$ obey
the equation \cite{tseytl}
\be
\ddot \lambda ^i (\tau ) + Q^2 \dot \lambda ^i (\tau )=
-Q^2\beta ^i (\lambda ) = -Q^2{\dot \lambda}^i =
-Q^2 G^{ij}\frac{\partial}{\partial \lambda ^j}
C (\lambda )
\label{eikosidyo}
\ee
Not only is this equation characteristic
of frictional motion in a potential $C(\lambda ^i )$,
(\ref{dodeka}),
but it is similar to the inflaton equation
(\ref{ena},\ref{dyo}).
Upon the inverse
rescaling $(\ref{resc})$, the friction term
becomes $\propto Q$, which implies that in standard
units the Hubble constant
\be
   H =\frac{Q}{3}
\label{eikositria}
\ee
The pioneering cosmological solution of \cite{aben}
can be regarded as a special case of (\ref{eikosidyo}),
in which the running of $Q$ was not considered.
As was pointed out in \cite{aben}, the appearance
of a central charge deficit $Q^2$ is accompanied by a
non-trivial dilaton potential, whose magnitude
determines the Hubble expansion rate.
\pr
Our discussion of a string inflationary scenario requires
the appropriate generalization\cite{emnerice}  of the
dissipative equation (\ref{eikosidyo}) to the statistical
(equilibrium) distribution function
in string theory space
$ \rho (\lambda ^i, p_j, \tau)$.
The $p_i$ denote momenta conjugate
to the `coordinates'
$\lambda ^i$, whose expectation values
are identified with the pertinent
vertex operators $V_i$.
The renormalizability of the world-sheet $\sigma$-model
tells us that
\bea
 {\cal D} S_{mL} &=& 0 \nn \\
{\cal D} \equiv
\frac{d}{d\tau }  =  \partial _\tau +
{\dot \lambda}^i \partial _i + {\dot p}_i \partial _{p_i} &=&
\partial_\tau +
\beta ^i \partial _i + {\dot p}_i \partial _{p_i}
\label{eikosipente}
\eea
and in particular
that
\be
   {\cal D} \rho = 0 = \frac{\partial}{\partial \tau } \rho
+ \beta ^i \frac{\partial}{\partial \lambda ^i}
\rho + {\dot p}^i \frac{\partial}{\partial p_i}
\rho
\label{eikosiexi}
\ee
Substituting in (\ref{eikosiexi})
using the classical equations of motion
derivable from (\ref{ennea}) :
\be
{\dot \lambda}^i = \beta ^i = \frac{\partial}{\partial p_i}
{\cal
H}
\qquad ; \qquad {\dot p}_i = - \frac{\partial }{\partial \lambda^i}
{\cal H} - G_{ij}\beta ^j
\label{eikosiepta}
\ee
we derive\cite{emnerice}
\be
\frac{\partial }{\partial \tau} \rho
= - \{ \rho, {\cal H} \} - \beta ^i G_{ij} \frac{\partial \rho }{\partial
p_j}
\label{eikosiokto}
\ee
where $\{ , \}$ denotes the conventional Poisson bracket,
reflecting the fact that we are still working
at the classical level.
The second term in (\ref{eikosiokto})
represents dissipation\cite{santilli}
due to couplings
with unobservable modes.
\pr
\section{Quantum Consistency}
\pr
As is well known, higher-genus effects in string theory
impose quantization of the effective string
couplings $\lambda^i$ \cite{strings}.
Since our
goal is a full quantum description of the string inflation
problem, we must check that the equations (\ref{eikosidyo})
are consistent with canonical quantization:
\be
[ \lambda ^i, \lambda ^j ] = 0
\qquad ; \qquad  [ p_i, p_j ] =0 \qquad ; \qquad
[\lambda ^i, p_j ] = -i \hbar \delta _j^i
\label{triantaena}
\ee
This is true if there is an underlying Lagrangian $L$ \cite{hojman}
whose equations of motion are equivalent but not necessarily
identical to (\ref{eikosidyo}), i.e. if there exists
a non-singular matrix $w_{ij}$ :
\be
w_{ij} ({\ddot \lambda} ^j + 2Q^2 {\dot \lambda}^j ) =
\frac{d}{d \tau }(\frac{\partial L}{\partial {\dot \lambda} ^i}   )-
\frac{\partial L}{\partial \lambda ^i}
\label{triantadyo}
\ee
which obeys
the following Helmholtz conditions :
\bea
w_{ij} &=& w_{ji} \nn \\
\frac{\partial w_{ij}}{\partial {\dot \lambda}^k} &=&
\frac{\partial w_{ik}}{\partial {\dot \lambda}^j}
\nn \\
   \frac{1}{2}\frac{D}{D \tau }
(w_{ik}\frac{\partial
f^k}{\partial {\dot \lambda}^j}
- w_{jk}\frac{\partial
f^k}{\partial {\dot \lambda}^i})&=&
w_{ik} \frac{\partial f^k}{\partial
\lambda ^j} -
w_{jk} \frac{\partial f^k}{\partial
\lambda ^i}\nn \\
\frac{D}{D \tau} w_{ij} &=&
-\frac{1}{2} w_{ik}\frac{\partial
f^k}{\partial {\dot \lambda}^j}
-\frac{1}{2} w_{jk}\frac{\partial
f^k}{\partial {\dot \lambda}^i}
\label{33}
\eea
where
\be
f^i \equiv -2Q^2 {\dot \lambda }^i
\qquad ; \qquad
\frac{D}{D \tau } \equiv \partial _\tau + {\dot \lambda ^i} \partial _i
+ f^i \frac{\partial }{\partial {\dot \lambda}^i }
\label{34}
\ee
If the conditions (\ref{33}) are met, then
\cite{hojman}
\be
  w_{ij}  =  \frac{\partial ^2 L } {\partial {\dot \lambda ^i}
\partial {\dot \lambda ^j}}
\label{35}
\ee
The Lagrangian in (\ref{35}) can be determined up to
total derivatives according to \cite{santilli,hojman}:
\be
    {\cal S}  \equiv \int d\tau L =
    - \int d\tau \int _{0}^1 d\kappa \lambda ^i
    E_i (\tau, \kappa \lambda, \kappa {\dot \lambda},
\kappa {\ddot \lambda} )~;~
 E_i (\tau,\lambda,{\dot \lambda},{\ddot \lambda})  \equiv
  w_{ij}( {\ddot \lambda}^j
 + Q^2{\dot \lambda}^i + Q^2\beta ^i )
\label{36b}
\ee
In our case,
near a fixed point
where the variation in $Q$ can be neglected,
up to $O[\beta^3]$ it acquires the form
\be
{\cal S}
= -\frac{1}{2}\int d\tau ( {\dot \lambda}^i G_{ij}[\lambda,\tau]
  {\dot \lambda}^j
 + \dots )
\label{36c}
\ee
with the $\dots $ denoting
terms that can be removed by a renormalization scheme change.
Within a critical-string (on-shell) approach,
the action (\ref{36b},\ref{36c}) can be
considered as an effective
action generating the string scattering amplitudes.
Here it should be considered
as a target-space `off-shell' action \cite{mm2}
for non-critical strings.
Comparing (\ref{36c}) with equation (\ref{35}),
and taking into account the renormalization group
invariance of $L$,
we see immediately
that one should identify
$    w_{ij} = - G_{ij} $,
the metric in coupling space.
\pr
We know that $G_{ij}$ is symmetric, so the
first of the Helmholtz conditions (\ref{33})
is satisfied. The next two conditions hold
automatically because of the gradient flow
property (\ref{dodeka}) and the fact that
$G_{ij}$ and $C$ (\ref{ennea})
are functions of the coordinates $\lambda ^i$
and not of the conjugate momenta. Finally,
the fourth Helmholtz condition provides the
condition
\be
  \frac{D}{D \tau} G_{ij} = 2Q^2 G_{ij}
\label{37}
\ee
which implies
an expanding scale factor for the metric in string theory
space
\be
    G_{ij} [\tau,
    \lambda (\tau) ]= e^{2Q^2\tau} {\hat G}_{ij} [\tau ,
    \lambda (\tau) ]
\label{38}
\ee
where ${\hat G}_{ij}$
is a Liouville-renormalization-group invariant
function. This is exactly the form
of the Zamolodchikov metric in Liouville strings \cite{emnestimates}.
Thus there is indeed
an underlying Lagrangian, one can quantize consistently,
and
\be
 \partial _\tau \rho = i [\rho, {\cal H}]
 + i \beta ^j G_{ji} [\lambda ^i, \rho ]
\label{39}
\ee
is the appropriate quantum version of the
density matrix equation (\ref{eikosiokto}).
\pr
It is straightforward that the observable
energy $E = << {\cal H} >> = Tr(\rho {\cal H}) $ is conserved in
this approach\cite{emncpt}, as
a consequence of the renormalizability
of the underlying two-dimensional field-theoretic
formalism,
\bea
\partial _\tau <<{\cal H}>> &=& <<\partial_\tau C +
\partial _\tau {\cal H} >> = \nn \\
=\partial _\tau <<C>>  &+&
<< \partial _\tau {\cal H} >>
\propto
\partial _\tau << \partial _i \beta ^j >> =0
\label{energcons}
\eea
where one took into account that $C$ is a
renormalization- group-invariant functional
of $\lambda ^i$ only.
However, the quantum energy fluctuations
$\delta E~\equiv~[<<{\cal H}^2>> - (<<{\cal H}>>)^2 ]^{\frac{1}{2}}$
are time-dependent :
\be
\partial _\tau (\delta E)^2
= -i <<[\beta ^i , {\cal H}]\beta^j G_{ji}>> =
<<\beta^j G_{ji} \frac{d}{d \tau}\beta ^i >>
\label{40}
\ee
Using the fact that $\beta ^i G_{ij} \beta ^j $ is a
renormalization-group invariant quantity,
we can express (\ref{40}) in the form
\be
  \partial _\tau (\delta E)^2  = -
<<Q^2  \beta ^i G_{ij} \beta ^j >>= -
<<Q^2  \partial _\tau C >>
\label{41}
\ee
where equation (\ref{37}) has been taken into account.
This is the basis for the estimate we present later
of the quantum density fluctuations generated
during the string inflationary epoch.
\pr
\section{String Fokker-Planck Equation}
\pr
Having established  the canonical quantization
of the dynamics in string theory space,
we now sketch the derivation of the string
Fokker-Planck equation that incorporates
quantum effects in the renormalization
group flow equation (\ref{eikosidyo}).
We interpret this as an equation for the
classical coupling $\lambda _c (\tau)$,
and decompose the value $\lambda _c (\tau + \delta \tau)$
of the coupling at an infinitesimally
later time as
\be
 \lambda^i _c (\tau + \delta \tau) =
 \lambda^i_c(\tau) + {\dot \lambda}_c^i (\tau)\delta \tau
  + \delta \lambda_q^i
(\delta \tau )
\label{fp1}
\ee
where $\delta \lambda _q (\tau)$ is the quantum fluctuation.
We assume that this has a characteristic white-noise
distribution
\be
   P(\delta \lambda_q)=
[  \frac{1}{2\pi <\delta \lambda_q>^2}]^{\frac{1}{2}}
exp(-\frac{(\delta \lambda_q  )^2}{2<(\delta \lambda_q)^2>} )
\label{fp2}
\ee
arising from unobservable global string modes
that extend beyond the horizon.
This assumption can be justified in standard field-theoretic
inflation by an analysis of quantum fluctuations
in de Sitter space \cite{vilenkin,linde}.
The analogue in our case
is
\be
    <(\delta \lambda_q )^2>=\frac{Q^6}{4\pi^2}\delta \tau
\label{fp3}
\ee
which we will justify further {\it a posteriori}.
White noise in renormalization
group flow has been assumed previously \cite{mandal},
by analogy with the conventional Langevin equation
(gradient  flow form of the $\beta$-functions \cite{zam}).
In our interpretation this random noise
is induced by the sum over unobservable
string modes beyond the event horizon,
whose effects are captured in our approach by the
Liouville time field and its quantum fluctuations.
\pr
It is straightforward to derive
the string Fokker-Planck equation from
(\ref{eikosidyo},\ref{fp1},\ref{fp2},\ref{fp3}). The
probability distribution ${\cal P}(\lambda , \tau)$
in string coupling space may be represented as
as
\be
{\cal P}(\lambda, \tau) = \int d\lambda _c P(\delta \lambda _q)
{\cal P}
(\lambda _c, \tau)
\label{fp4}
\ee
where the integral is over all classical variables
$\lambda _c$ whose evolution in time yields the variable
$\lambda $ at rescaled time $\tau$.
Changing variables to the quantum fluctuations
$\delta \lambda _q$, we find
\be
  {\cal P}(\lambda, \tau)=
  \int d(\delta \lambda_q) {\cal P}
  (\lambda^i-\partial ^i C(\lambda)
  \delta \tau - \delta \lambda^i_q )
  det(1 + \frac{\partial }{\partial \lambda}
(\partial _\lambda C(\lambda))-
\partial _\lambda (\delta \lambda _q) )
\label{fp5}
\ee
It is a simple matter to make a Taylor
expansion \cite{linde} of the integrand
in (\ref{fp5}) and perform the Gaussian
integral over $\delta \lambda _q $ to obtain
the string Fokker-Planck equation
\be
\partial_\tau {\cal P}(\lambda, \tau) =
\frac{1}{8\pi^2}
\frac{\delta}{\delta \lambda ^i}
Q^{3}
\delta^{ij}
\frac{\delta}{\delta \lambda ^j}[Q^{3}
{\cal P} (\lambda,\tau)]
+ \frac{\delta }{\delta \lambda ^i}
[\beta ^i
{\cal P} (\lambda,\tau)]
\label{fp6}
\ee
modulo ordering ambiguities
for the $\lambda$-dependent diffusion coefficients.
The reader may verify that the dispersion $<(\delta \lambda _q )^2>$
calculated from (\ref{fp6}) is of the assumed form
(\ref{fp3})\footnote{Energy conservation (\ref{energcons})
can also be derived using the Fokker-Planck equation,
when one takes into account restrictions on the
probability distributions that are consequences of the
renormalization group.}.
Equation (\ref{fp6}) may alternatively be derived \cite{emnestimates}
from equation
(\ref{39}) for the statistical distribution
in string theory space, $\{ \lambda ^i \}$, if one assumes
\be
  \rho \simeq e^{-{\overline {\cal H}}}
  W(\lambda^i, \tau)
\label{fp7}
\ee
as in the conventional \cite{davidoff} treatment of Markov
processes, where
${\overline {\cal H}} \equiv -\frac{8\pi^2{\cal H}}{Q^6}$
with $Q^6/8\pi^2$
playing the r\^ole of an
effective `temperature' in theory space.
\pr
The string Fokker-Planck equation (\ref{fp6})
can be used to demonstrate that the probability
distribution function ${\cal P}$ takes the Gibbs
form \cite{fokker}
$ {\cal P} = e^{-{\overline {\cal H}}} $
asymptotically at large
times, i.e. that $W(\lambda ,\tau) \rightarrow 1$ as $\tau \rightarrow
\infty$. This is done by first noting
that the non-self-adjoint first-derivative term
in (\ref{fp6})
can be written in the form
\be
   \frac{\partial }{\partial \lambda^i} ( \beta ^i
   {\cal P}(\lambda ,\tau))
   = -\frac{\partial }{\partial \lambda ^i} ( G^{ij}\frac{\partial
   {\cal H}}
 {\partial \lambda ^j}  {\cal P}(\lambda , \tau))
\label{fp8}
\ee
using the fact that ${\dot p}_i =0$ in our scheme.
The quantity ${\hat {\cal P}} (\lambda ^i,\tau)
\equiv e^{\frac{1}{2}{\overline {\cal H}}}{\cal P}(\lambda ^i, \tau)$
therefore obeys the equation
\be
 {\hat O}_{FP} {\hat {\cal P}}(\lambda ^i, \tau)  =  0 ; \qquad
{\hat O}_{FP}  \equiv
-\frac{Q^6}{8\pi^2}[-\frac{\partial}
{\partial \lambda ^i}
+ \frac{4\pi^2}{Q^6}\frac{\partial
{\cal H}}{\partial \lambda ^i}][\frac{\partial}
{\partial \lambda ^i} + \frac{4\pi^2}{Q^6}
\frac{\partial {\cal H}}{\partial \lambda ^i} ]
\label{fp9}
\ee
where ${\hat O}_{FP}$ is self-adjoint. Thus we can expand
\be
{\cal P}(\lambda ^i, t) = e^{-\frac{1}{2}{\overline {\cal H}}}
\sum _{n} a_n e^{\varepsilon _n t}
f_n (\lambda ^i )
\label{fp10}
\ee
where the $f_n (\lambda ^i)$ are eigenfunctions of ${\hat O}_{FP}$
with negative semi-definite eigenvalues $\varepsilon _n $:
\be
 \varepsilon _n = \int d\lambda ^i f^{*}_n {\hat O}_{FP} f_n =
 -\frac{Q^6}{8\pi^2}
 \int d\lambda ^i | (\frac{\partial}{\partial \lambda ^i}
- \frac{4\pi^2}{Q^6}
\frac{\partial {\cal H}}{\partial \lambda ^i})f_n |^2\le 0
\label{fp11}
\ee
where $\varepsilon _0 = 0$ for $f _0 =  e^{-\frac{1}{2}{\overline
{\cal H}}}$, and
the contraction of the indices is done with the
help of the positive definite Zamolodchikov metric $G_{ij}$.
It follows that ${\cal P}(\lambda ^i, t) \rightarrow
a_0e^{-{\overline {\cal H}}}$,
the equilibrium Gibbs distribution, at large times \footnote{This
result has been derived in a different way in
\cite{emnroma}, using pointer states
in string theory space.}.
\pr
\section{Realization of Inflation}
\pr
The string Fokker-Planck equation (\ref{fp6})
can be used to discuss the circumstances
under which the entropy
\be
 S=-k_B \int [D\lambda^i] {\cal P}(\lambda, \tau)
 ln {\cal P}(\lambda,\tau)
\label{I1}
\ee
may grow exponentially, as in inflationary cosmology.
Using (\ref{fp6}), we find
\be
\partial _\tau S = \int [D\lambda ^i] \frac{k_B}{2} [ \frac{
Q^6}{4\pi^2}
\frac{|\nabla _i {\cal P}|^2}
{{\cal P}} +
\frac{1}{8\pi^2}(\nabla _i Q^6)\nabla ^i {\cal P}  +
2\beta ^i \nabla _i {\cal P} ]
\label{I2}
\ee
As a result of the renormalization-group invariance
of ${\cal P}$,
the last term on the right-hand side of (\ref{I2})
contributes
boundary terms in string theory space, which we
discard in our cosmological context.
We assume that
the second term
$\propto \nabla ^i Q^6$
is much smaller than
the first term, corresponding to an inflationary
epoch during which
$Q$ is approximately constant, obtaining
\be
\partial _\tau S = \int [D\lambda ^i] \frac{k_B}{8\pi^2} [ Q^6
\frac{|\nabla _i {\cal P}|^2}
{{\cal P}}] \ge 0
\label{I3}
\ee
The positivity of (\ref{I3}) is a consequence
of world-sheet unitarity, and the zero occurs only
at fixed points where $Q \rightarrow 0$. Using now
the Gibbs formula
which implies that $ln{\cal P} = -{\overline {\cal H}} =
\frac{8\pi^2{\cal H}}{Q^6}$,
and recalling that $\nabla ^i {\cal H} \simeq -\beta ^i $,
we see that
\be
{\dot S} = -k_B \int [D\lambda ] \frac{1}{|{\cal H}|}
(\beta ^i \beta _i )
{\cal P} ln {\cal P}
\label{I4}
\ee
It is apparent that if $\beta ^i \beta _i \simeq Q^4 $
varies sufficiently slowly over a sufficiently long
period of time - the string analogue of the
``slow rollover '' assumption of conventional inflation
\cite{vilenkin,linde} -
the entropy will indeed grow exponentially :
\be
   S \propto e^{Q^2\tau} = e^{Qt}
\label{I4b}
\ee
since ${\cal H}$ is of $O[Q^2]$ during inflation.
The re-expression of (\ref{I4b}) in terms
of the physical time $t$
confirms the correspondence (\ref{eikositria})
of the central charge deficit $Q$
to the conventional Hubble expansion rate during inflation.
\pr
This growth in the entropy is accompanied
by growth in the scale factor of the Universe,
as can be seen by considering the time-dependence
of the space-time volume element $V$:
\be
    \partial _\tau V =\partial _\tau (\int d^DX  \sqrt{G}e^{-\Phi})
=\partial _\tau (e^{Q^2\tau}
\int d^DX\sqrt{G})=(Q^2 +  \partial _\tau (Q^2) ) V
\label{V1}
\ee
The second term in (\ref{V1}) is
$\propto O[(\beta ^i )^2 ]$, and can be removed
in an appropriate renormalization scheme. For (approximately)
constant $Q$, the volume element (\ref{V1}) indeed grows
(approximately) exponentially : $V \propto e^{Qt}$.
\pr
In this scenario, inflation occurs without
an associated inflaton field, whose r\^ole
is taken over by the Liouville time variable.
Inflation is a quantum effect, as in Starobinsky's
proposal \cite{staro}. However, our scenario is formulated
in terms of quantum fluctuations on the
world sheet.
\pr
To gain some impression how this scenario
for string inflation might work
in a specific model, consider the
$SL(2,R)/U(1)$ model of a stringy black hole \cite{witt}
with central charge
$c=\frac{3k}{k -1} - 1$, for which the
Friedman-Robertson-Walker line element
\be
  ds^2 = dt^2 - R(t)^2 dr^2 \qquad : \qquad
R^2 (t)= k(t) -2
\label{V2}
\ee
where the renormalized level parameter $k(t)$
is allowed to run as in (\ref{eikosidyo}).
It is easy to check that $R(t)$ exhibits
an exponentially-growing Jeans instability \cite{emnharc,jeans}
if
\be
   2 Q (k(t)-2) >  \frac{k - 1}{k}
  |{\dot k}(t)|
\label{V3}
\ee
which is the analogue of the slow rollover condition
in this model.
\pr
We conclude with a discussion
of density fluctuations
in our scenario. Using (\ref{dekapente}) in the
estimate (\ref{41}), we find
\be
   \partial _\tau (\delta E)^2 = - <<[(\frac{1}{3}
\delta C + \frac{1}{4} \beta ^i G_{ij} \beta ^j )
\partial _\tau \delta C ] >>
=
- <<[(\frac{1}{3}
\delta C + \frac{1}{4} Q^2 + \frac{1}{4} (({\dot \lambda }')^i )^2)
\partial _\tau \delta C ] >>
\label{V4}
\ee
where $\delta C = C - 25$,
the $Q^2$ term on the right-hand-side
is associated with graviton fluctuations
in inflationary
backgrounds (\ref{V1}), upon taking
into account some properties
\cite{emnestimates}
of the
Quantum Zamolodchikov metric $G_{ij}$
in theory space
in the class of renormalization
schemes appropriate for inflation,
which imply  that
$<<\beta ^i G_{ij} \beta ^j>>=Q^2$.
The
$(\lambda ')^i$ in (\ref{V4})
denote the remaining
matter fields.
Using the fact that $({\dot \lambda}')^i$,
being a $\beta$-function of a renormalizable coupling,
does not depend explicitly on the scale $\tau$,
we
may
cast (\ref{V4})
in the form
\be
\partial _\tau (\delta E)^2 = <<
-2\partial _\tau [ \frac{1}{3}\delta C +
\frac{1}{4}(({\dot \lambda}')^i)^2 ]^2
+ \frac{1}{8}\partial _\tau (({\dot \lambda}')^i )^4 >>
\label{F2}
\ee
The first term in the right-hand side of (\ref{F2})
is of higher order in the $\beta ^i$
than the second term. This follows from an
analysis\cite{emnestimates}
of the temporal evolution of the
string density, which
is constant in time during inflation \cite{turok}.
Hence we estimate:
\be
(\delta E) ^2 \simeq
\frac{1}{8} <<(({\dot \lambda ')}^i)^4 >> + \dots
\label{F3}
\ee
where  the $\dots $ represent
higher-order terms originating during our
string inflationary epoch, as well as
any other possible contributions
to $(\delta E)^2$.
\pr
We identify $(\delta E)^2$
with the usual r.m.s. energy-density
fluctuation $(\delta \rho )^2$.
To estimate $\frac{\delta \rho }{\rho }$
as observed, for example, by COBE \cite{COBE},
we recall\cite{guth,vilenkin} that
the quantity $\delta \rho / (\rho + p) $,
with $p$ the pressure,
is gauge-invariant and remains constant
while it is beyond the horizon. In our approach
\be
  \rho + p \simeq <<({\dot \lambda}^i)^2>>
  \simeq  Q^2 +  <<( ({\dot \lambda}')^i)^2  >>
\label{F4}
\ee
during inflation, and we expect $Q^2 >>
<<(({\dot \lambda} ')^i)^2 >> $
as our formulation
of the conventional slow roll-over condition \cite{guth,vilenkin},
leading to the estimate :
\be
  ( \frac{\delta \rho }{\rho })_{COBE} \simeq
\frac{1}{2\sqrt{2}}\frac{<<(({\dot \lambda '}))^2>>}{Q^2}
=
\frac{1}{2\sqrt{2}}<<(\frac{ d(\lambda ')^i}{d t})^2>>
\label{F5}
\ee
Formally we expect
$<<[\frac{d (\lambda ')^i}{dt}]^2>> \simeq O[M_s ^4]$,
where $M_s$ is the string unification scale
\cite{lacaze}: $M_s~\le~\frac{1}{10} $ Planck units.
We therefore expect
\be
    (\frac{\delta \rho }{\rho })_{COBE}
\le 10^{-4}
\label{F6}
\ee
but are not yet in a position
to make more detailed estimates, which would
require an exact model of the particle theory
effective below the Planck scale.
\pr
\pr
\pr
\nk {\Large{\bf  Acknowledgements}}
\pr
The work of N.E.M. is supported by a EC Research Fellowship,
Proposal Nr. ERB4001GT922259 .
That of D.V.N. is partially supported by DOE grant
DE-FG05-91-GR-40633.
\vfil\eject

\end{document}